\documentclass[conference]{IEEEtran}
\usepackage{amsmath}
\usepackage{cite}
\usepackage{graphicx}
\usepackage{url}
\usepackage{bbm}
\usepackage{stfloats}
\usepackage{epstopdf}

\usepackage{txfonts}
\usepackage[thinlines]{easytable}

\DeclareMathOperator{\dB}{dB}
\DeclareMathOperator{\bps}{bps}
\DeclareMathOperator*{\st}{s.t.}

\newtheorem{theorem}{\textbf{Theorem}}

\newtheorem{lemma}{\textbf{Lemma}}

\begin{document}


\title{Short-Packet Communications in Non-Orthogonal Multiple Access Systems}

\author{\IEEEauthorblockN{
		Xiaofang Sun\IEEEauthorrefmark{1}\IEEEauthorrefmark{2},
		Shihao Yan\IEEEauthorrefmark{2},
		Nan Yang\IEEEauthorrefmark{2},
		Zhiguo Ding\IEEEauthorrefmark{3},
		Chao Shen\IEEEauthorrefmark{1}, and
		Zhangdui Zhong\IEEEauthorrefmark{1}\IEEEauthorrefmark{4}}

\IEEEauthorblockA{\IEEEauthorrefmark{1}State Key Lab of Rail Traffic Control and Safety, Beijing Jiaotong University, Beijing, China\\
\IEEEauthorrefmark{2}Research School of Engineering, The Australian National University, Canberra, Australia\\
\IEEEauthorrefmark{3}School of Computing and Communications, Lancaster University, Lancaster, UK\\
\IEEEauthorrefmark{4}Beijing Engineering Research Center of High-speed Railway Broadband Mobile Communications}
\IEEEauthorblockA{Email: \{xiaofangsun, chaoshen, zhdzhong\}@bjtu.edu.cn,
\{shihao.yan, nan.yang\}@anu.edu.au, z.ding@lancaster.ac.uk}\vspace{-4mm}}

\markboth{Submitted to IEEE GLOBECOM 2017}{Sun \MakeLowercase{\textit{et al.}}: Short-Packet Communications in Non-Orthogonal Multiple Access Systems}

\maketitle
\begin{abstract}
This work introduces, for the first time, non-orthogonal multiple access (NOMA) into short-packet communications to achieve low latency in wireless networks. Specifically, we address the optimization of transmission rates and power allocation to maximize the effective throughput of the user with a higher channel gain while guaranteeing the other user achieving a certain level of effective throughput. To demonstrate the benefits of NOMA, we analyze the performance of orthogonal multiple access (OMA) as a benchmark. Our examination shows that NOMA can significantly outperform OMA by achieving a higher effective throughput with the same latency or incurring a lower latency to achieve the same effective throughput targets. Surprisingly, we find that the performance gap between NOMA and OMA becomes more prominent when the effective throughput targets at the two users become closer to each other. This demonstrates that NOMA can significantly reduce the latency in the context of short-packet communications with practical constraints.
\end{abstract}

\IEEEpeerreviewmaketitle

\section{Introduction}

In the fifth generation (5G) wireless ecosystem, the majority of wireless connections will most likely be originated by autonomous machines and devices~\cite{Durisi_2016}. Against this background, machine-type communications (MTC) are emerging to support complementary services to that provided by mobile broadband networks. Specifically, massive MTC and ultra-reliable MTC have been identified as two key application scenarios of 5G wireless networks~\cite{METIS}. In MTC, the requirement on latency (besides the reliability) is stringent (e.g., 4 ms)~\cite{Durisi_2016}, since low latency is pivotal to ensure real-time functionality in interactive communications of machines. In some industrial automation applications, for example, short packets consisting of approximately 100 bits are desired to be transmitted within 100 $\mu$s~\cite{Johansson_2015}. The challenge in achieving such a low latency lies in the capability to support short-packet communications.

In short-packet communications, as pointed out by~\cite{FBL_2010}, the decoding error probability at a receiver is not negligible since the length of a codeword is finite (i.e., the blocklength is finite). This is different from the Shannon capacity theorem, in which the decoding error probability is negligible as the blocklength approaches infinity. Considering the effect of decoding errors, the channel coding rate in finite blocklength regime was derived in
\cite{FBL_2010}. This pioneering work serves as the foundation in examining the performance of short-packet communications. Triggered by~\cite{FBL_2010}, the impact of finite blocklength on different communication systems has been widely studied. For example,  the achievable channel coding rate in quasi-static multiple-input multiple-output (MIMO) fading channels was examined in~\cite{Yang_2014}. In~\cite{Durisi_2016_TCOM}, the tradeoff between reliability, throughput, and latency in short-packet communications was investigated over Rayleigh fading channels. Furthermore, the information theoretic result in \cite{FBL_2010} was utilized for packet scheduling of a multi-user scenario, wherein the latency-critical packets are transmitted in orthogonal channels \cite{Xu_2016}.

Recently, non-orthogonal multiple access (NOMA) has attracted increasing research interests since it has been recognized as a promising technique that provides superior spectrum efficiency in 5G wireless networks~\cite{Ding_CM_2017}. NOMA is a multi-user multiplexing scheme that achieves multiple access in the power domain~\cite{Saito_2013}. The benefits of NOMA have been widely examined in various wireless communications, such as broadcast channels~\cite{Sun_2016,Ding_2016_TWC}, full-duplex communications~\cite{FD_2017}, and physical layer security~\cite{Qin_2016_ICC,PLS_2017}. However, its potential benefits in latency reduction in the context of short-packet communications have never been revealed. Turning the question around, the impact of finite blocklength on the performance of NOMA has never been examined, although this impact has been widely investigated for other communications paradigms, as discussed in the previous paragraph. This leaves an important gap in our understanding on the benefits of NOMA in the context of short-packet communications and the impact of finite blocklength on NOMA, which motivates this work.




In this work, \emph{for the first time} we introduce NOMA into short-packet communications in which the benefits of NOMA in latency reduction are explored. We consider a scenario where a single-antenna base station (BS) serves two single-antenna users in the downlink broadcast channel. Specifically, we determine the optimization of transmission rates and power allocation to maximize the effective throughput of one user subject to the constraint on the effective throughput of the other user. In order to explicitly demonstrate the benefits of NOMA, we analyze the performance of orthogonal multiple access (OMA) in the context of short-packet communications as a benchmark. Our examination indicates that NOMA can significantly outperform OMA by achieving a higher effective throughput at one user (subject to the same constraint on the effective throughput at the other user) with the same latency or incurring a lower latency to achieve the same effective throughput targets. Surprisingly, our results show that the performance gap between NOMA and OMA becomes more prominent when the desired effective throughput targets at the two users become closer to each other.

%
%
%



\section{System Model}\label{sec:system_model}

\begin{figure}[!t]
\centering
    \includegraphics[width=0.7\columnwidth]{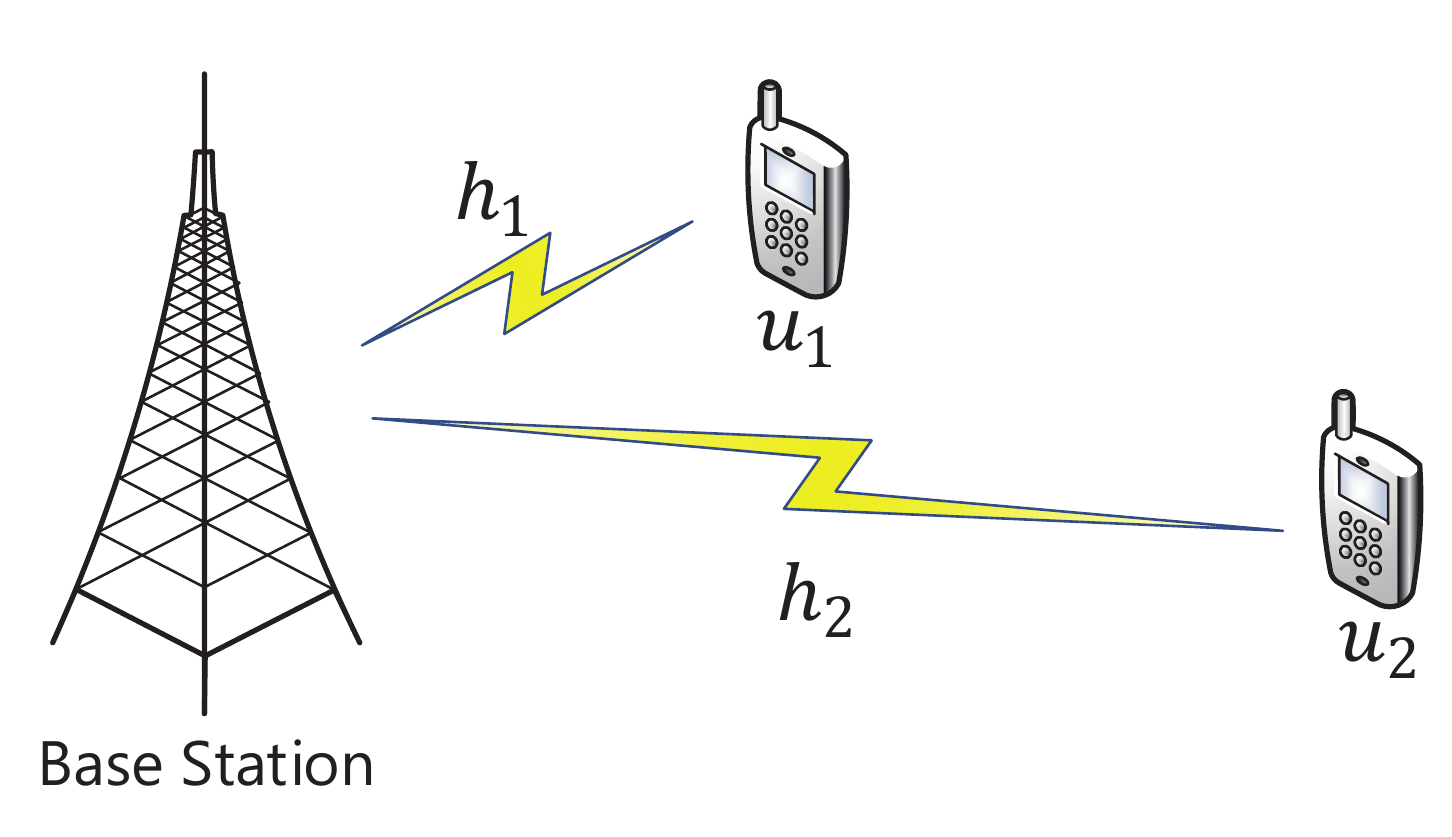}
    \caption{Illustration of a NOMA system where a single-antenna BS communicates with two single-antenna users.}\label{fig:system_model}\vspace{-4mm}
\end{figure}


In this work, we consider a downlink broadcast channel, as depicted in Fig.~\ref{fig:system_model}, in which a BS serves two single-antenna users at the same frequency within a finite blocklength. As only one antenna is equipped at BS, it serves the two users either over different time slots (OMA) or simultaneously by adopting NOMA. We define the user with a higher channel gain as user 1 (denoted by $u_1$) and the other one as user 2 (denoted by $u_2$). The channel coefficients between the BS and $u_1$ and between the BS and $u_2$ are denoted by $h_1$ and $h_2$, respectively. We assume that $h_1$ and $h_2$ are subject to independent quasi-static Rayleigh fading with equal block length and the associated channel gains are perfectly known by BS. Given the channel gain relationship between $u_1$ and $u_2$, we have $\left|h_1\right|\geq\left|h_2\right|$.


\subsection{Achievable Rate with Finite Blocklength}

As per Shannon's channel coding theorem, the decoding error probability at the receiver becomes negligible as the blocklength  $N$ approaches infinity~\cite{Shannon_1948}. Short-packet communications aims to achieve low latency (e.g., short delay), in which the blocklength should be finite and normally of a small value. As pointed out by~\cite{FBL_2010}, the decoding error probability at the receiver is nonnegligible when the blocklength $N$ is finite. Thus, we denote $R_{i}$ as the channel coding rate with finite blocklength $N_i$ for a given decoding error probability $\epsilon_i$ at user $i$ ($i = 1$, $2$), and express it as~\cite{Ozcan_2013,FBL_2010}
\begin{align}\label{eq:FBL_Rate}
R_i = \log_2(1+\gamma_i)-\sqrt{\frac{V_i}{N_i}}\frac{Q^{-1}(\epsilon_i)}{\ln 2},
\end{align}
where $\gamma_i$ denotes the signal-to-noise ratio (SNR) at $u_i$, $V_i$ is the channel dispersion given by
\begin{align}
V_i=1-\left(1+\gamma_i\right)^{-2},
\end{align}
${Q}^{-1}(\cdot)$ is the inverse Q-function. Equivalently, for a given channel coding rate $R_i$, the decoding error probability at $u_i$ is approximated by
\begin{align}\label{eq:EP}
\epsilon_i &\triangleq\mathcal{G}\left(\gamma_i, N_i, R_i\right)\notag\\
&=Q\left(\ln2\sqrt{\frac{N_i}{V_i}}\left(\log_2\left(1+\gamma_i\right)-R_i\right)\right).
\end{align}

In this paper, we adopt the effective throughput as the objective metric to evaluate the system performance with finite blocklength. Mathematically, the effective throughput ($\bps$) at $u_i$, $T_i$, is given by
\begin{align}\label{eq:Throughput}
T_i = \frac{N_i}{N} R_i (1 - \bar{\epsilon}_i),
\end{align}
where $\bar{\epsilon}_i$ is the effective decoding error probability at $u_i$.
For fixed $N_i$, there exists an optimal value of $R_i$ (or equivalently $\epsilon_i$) that maximizes the effective throughput $T_i$.


\subsection{Optimization Problem}

%


In our system, the BS needs to serve $u_{1}$ and $u_{2}$ within $N$ symbol periods (i.e., the finite blocklength is $N$). The ultimate goal of our design is to achieve the maximum effective throughput of $u_1$, defined as $T_1$, while guaranteeing some specific requirements on the effective throughput at $u_2$ subject to a total power constraint. Mathematically, the optimization problem for the BS with NOMA/OMA is formulated as
\begin{subequations}\label{eq:Opt}
\begin{align}
\underset{\Delta}{\max} ~~~ &T_1 \label{eq:Ojective}\\
\st~~~~&T_2 \geq \overline{T}_2,\label{eq:C_T}\\
& P_1N_1+P_2N_2\leq \overline{P}N,\label{eq:C_Energy}\\
& \Psi\left(N_1,N_2\right) \leq N,\label{eq:C_time}
\end{align}
\end{subequations}
where $\Delta=\{R_1, R_2, P_1, P_2, N_1, N_2\}$ represents the variable set that needs to be determined at BS, $\overline{T}_2$ is the minimum required effective throughput at $u_2$, $P_1$ and $P_2$ are the allocated transmit powers to $u_1$ and $u_2$, respectively, $\overline{P}$ is the average transmit power within one fading block, $\Psi\left(N_1,N_2\right)=\max\left\{N_1,N_2\right\}$ for NOMA, and $\Psi\left(N_1,N_2\right)=N_1+N_2$ for OMA. As per \eqref{eq:Opt}, we need to design the power allocation and time slots allocation at the BS on top of determining the transmission rates for $u_1$ and $u_2$, such that $T_1$ is maximized subject to given constraints.



\section{Design of Non-Orthogonal Multiple Access}\label{sec:NOMA}

In this section, we design a new NOMA scheme to solve the optimization problem given in \eqref{eq:Opt}. In the designed scheme, superposition coding (SC) is employed in the transmission such that BS is able to transmit to $u_1$ and $u_2$ simultaneously at different power levels. This implies the use of $N_1=N_2=N$ in the designed scheme.


\subsection{Transmission to User 1}

When the BS adopts NOMA to transmit, the received signal at $u_1$ at each symbol period is given by
\begin{align}\label{eq:y1}
y_1=h_1x+n_1=h_1(\sqrt{P_1}x_1+\sqrt{P_2}x_2)+n_1,
\end{align}
where $x=\sqrt{P_1}x_1+\sqrt{P_2}x_2$ is the transmitted signal at the BS, $x_1$ and $x_2$ represent the information signals to $u_1$ and $u_2$, respectively, and $n_1\sim\mathcal{CN}(0,\sigma_{1}^2)$ denotes the additive white Gaussian noise (AWGN) at $u_1$ with zero mean and variance $\sigma_{1}^2$. Since $\left|h_{1}\right| \geq \left|h_{2}\right|$, we consider that successive interference cancelation (SIC) is employed at $u_1$ to remove the interference caused by $x_2$. To do this, $u_1$ first decodes $x_2$ treating $x_1$ as the interference. Following \eqref{eq:y1}, the signal-to-noise-plus-interference ratio (SINR) of $x_2$ at $u_1$, denoted by $\gamma_{2}^{1}$, is given by
\begin{align}\label{eq:SINR_SIC}
\gamma_{2}^{1}=\frac{P_2\left|h_{1}\right|^2}{P_1\left|h_{1}\right|^2+\sigma_{1}^{2}}.
\end{align}
Following \eqref{eq:EP}, the decoding error probability of $x_2$ at $u_1$ for given $R_2$, denoted by $\epsilon_{2}^{1}$, is approximated by $\epsilon_{2}^{1} = \mathcal{G}(\gamma_{2}^{1}, N, R_2)$. Accordingly, the probability that $x_2$ is correctly decoded and completely canceled at $u_1$ is $1-\epsilon_{2}^{1}$. These indicate that perfect SIC may not be guaranteed in our system due to the use of finite blocklength, which is different from the case using infinite blocklength and achieving perfect SIC~\cite{Ding_CM_2017}.

If perfect SIC cannot be guaranteed (which occurs with the probability $\epsilon_{2}^{1}$), $u_1$ has to decode $x_1$ directly subject to the interference caused by $x_2$. Correspondingly, the SINR of $x_1$, denoted by $\gamma'_{1}$, is given by
\begin{align}\label{eq:SINR1}
\gamma'_{1}=\frac{P_1\left|h_{1}\right|^2}{P_2\left|h_{1}\right|^2+\sigma_{1}^{2}}.
\end{align}
Then, the decoding error probability of $x_1$ for given $R_1$, denoted by $\epsilon'_{1}$, is approximated by $\epsilon'_{1} = \mathcal{G}(\gamma'_{1}, N, R_1)$.

Differently, if perfect SIC can be guaranteed, following \eqref{eq:y1} the SNR of $x_1$ at $u_1$, denoted by $\gamma_1$, is given by
\begin{align}\label{eq:SNR1}
\gamma_1=\frac{P_1\left|h_{1}\right|^2}{\sigma_{1}^{2}}.
\end{align}
Accordingly, the decoding error probability of $x_1$ for given $R_1$, denoted by $\epsilon_1$, is approximated by $\epsilon_1 = \mathcal{G}(\gamma_1, N, R_1)$.


Considering both~\eqref{eq:SINR1} and~\eqref{eq:SNR1}, we obtain the effective decoding error probability of $x_1$ at $u_1$ as
\begin{align}\label{eq:Average_EP1}
\overline{\epsilon}_1=(1-\epsilon_{2}^{1})\epsilon_1+\epsilon_{2}^{1}\epsilon'_{1}.
\end{align}

\subsection{Transmission to User 2}

When the BS adopts NOMA, the received signal at $u_2$ is given by
\begin{align}\label{eq:y2}
y_2=h_2x+n_2=h_2(\sqrt{P_1}x_1+\sqrt{P_2}x_2)+n_2,
\end{align}
where $n_2\sim\mathcal{CN}(0,\sigma_{2}^2)$ denotes the AWGN at $u_2$ with zero mean and variance $\sigma_{2}^2$.

Due to $\left|h_{1}\right| \geq \left|h_{2}\right|$, SIC is not needed at $u_2$. As such, $u_2$ decodes its own signal directly with the interference caused by $x_1$. The SINR of $x_2$ at $u_2$, denoted by $\gamma_2$, is given by
\begin{align}\label{eq:SINR2}
\gamma_2=\frac{P_2\left|h_{2}\right|^2}{P_1\left|h_{2}\right|^2+\sigma_2^2}.
\end{align}
Accordingly, the decoding error probability of $x_{2}$ for given $R_{2}$, denoted by $\epsilon_2$, is approximated by $\epsilon_2 = \mathcal{G}(\gamma_2, N, R_2)$. Since there only exists one decoding strategy at $u_2$, the decoding error probability $\epsilon_2$ is actually the effective decoding error probability at $u_2$, i.e., $\overline{\epsilon}_2=\epsilon_2$.


\subsection{Optimization Problem in NOMA}\label{sec:NOMA_opt}
%

In NOMA transmission, we have $N_1 = N_2 = N$. Therefore, we rewrite the optimization problem given in \eqref{eq:Opt} as
\begin{subequations}\label{eq:Opt1}
\begin{align}
\underset{\Delta_n}{\max} ~~~&T_1=R_1\left[\left(1-\epsilon_1\right)\left(1-\epsilon_{2}^{1}\right)
+\left(1-\epsilon'_{1}\right)\epsilon_{2}^{1}\right] \label{eq:Objective1}\\
\st~~~~&T_2\geq \overline{T}_2,\label{eq:C_T_noma}\\
&P_1+P_2 \leq \overline{P}, \label{eq:C_Energy_NOMA}
\end{align}
\end{subequations}
where $\Delta_n=\{R_1, R_2, P_1, P_2\}$ is the variable set that needs to be determined at the BS in NOMA.

In order to solve \eqref{eq:Opt1}, we first prove that the equality in \eqref{eq:C_Energy_NOMA} is always achieved when $T_{1}$ is maximized, in the following lemma.
\begin{lemma}\label{lemma1}
The equality in the power constraint \eqref{eq:C_Energy_NOMA}, i.e., $P_1+P_2=\overline{P}$, is always guaranteed when the effective throughput $T_1$ is maximized subject to $T_2\geq \overline{T}_2$.
\begin{IEEEproof}
Please refer to Appendix~\ref{App:inequality}.
\end{IEEEproof}
\end{lemma}

From the proof of \textit{\textbf{Lemma}~\ref{lemma1}}, we see that $T_2 > \overline{T}_2$ may happen in NOMA. However, in the following we consider $T_2 = \overline{T}_2$ in NOMA to guarantee a reasonable comparison between NOMA and OMA, since OMA maximizes $T_1$ when $T_2 = \overline{T}_2$ (which will be discussed in the next section). Also, we consider that $R_2$ is determined through maximizing $T_2$ only, which again may not be optimal in NOMA since $T_1$ is also a function of $R_2$. We note that this consideration may lead to an unfair comparison between NOMA and OMA since the performance of NOMA is not optimized while that of OMA is optimized. Nevertheless, if we show that NOMA outperforms OMA under these considerations, we can make a conclusion that NOMA definitely outperforms OMA.

With these considerations and following \textit{\textbf{Lemma}~\ref{lemma1}}, the power allocation and $R_2$ in NOMA are determined such that $T_2 = \overline{T}_2$ is guaranteed with $R_2$ maximizing $T_2$ for any given $P_2$ under the constraint $P_1+P_2 \leq \overline{P}$. As such, the optimal values of $P_1$, $P_2$, and $R_2$, denoted by $P_1^{\ast}$, $P_2^{\ast}$, and $R_2^{\ast}$, respectively, can be determined efficiently as per \eqref{eq:SINR2} and $\epsilon_2$, which are independent of $R_1$. We next determine the optimal value of $R_1$ that maximizes $T_1$ subject to the given constraints. To this end, in the following theorem we prove that there always exists a certain value of $R_1$ that maximizes $T_1$ and determine this optimal value.
\begin{theorem}\label{Theorem1}
There always exists a certain coding rate $R_1$ that maximizes the effective throughput of $u_1$. This optimal value, denoted by $R_{1}^{\ast}$, is the unique solution to
\begin{align}\label{eq:R1*}
&(1-\epsilon_{2}^{1})\left(1-Q\left(\frac{a-R_{1}^{\ast}}{b}\right)-\frac{R_{1}^{\ast}}{b\sqrt{2\pi}}e^{-\frac{(a-R_{1}^{\ast})^2}{2b^2}}\right)\notag\\
&=-\epsilon_{2}^{1}\left(1-Q\left(\frac{a'-R_{1}^{\ast}}{b'}\right)-\frac{R_{1}^{\ast}}{b'\sqrt{2\pi}}e^{-\frac{(a'-R_{1}^{\ast})^2}{2b'^2}}\right),
\end{align}
where $a=\log_2(1+\gamma_1)$, $a'=\log_2(1+\gamma'_{1})$,
\begin{align}
b=\frac{1}{\ln2}\sqrt{\frac{1}{N}\left(1-\frac{1}{(1+\gamma_1)^2}\right)},\notag
\end{align}
and
\begin{align}
b'=\frac{1}{\ln2}\sqrt{\frac{1}{N}\left(1-\frac{1}{(1+\gamma'_{1})^2}\right)}.\notag
\end{align}
\begin{IEEEproof}
Please refer to Appendix~\ref{App:NOMA}.
\end{IEEEproof}
\end{theorem}

Although $R_1^{\ast}$ cannot be derived in closed form, it can be determined efficiently due to the concavity. Substituting the determined $P_1^{\ast}$, $P_2^{\ast}$, $R_1^{\ast}$, and $R_2^{\ast}$ into \eqref{eq:Throughput}, we obtain the maximum $T_1$ achieved by NOMA, which is denoted by $T_1^{\ast}$.


\section{Design of Orthogonal Multiple Access}\label{sec:OMA}

In this section, we adopt OMA in the system and solve the optimization problem given in \eqref{eq:Opt}. In OMA, the two users are served in orthogonal time slots and thus, $N_1+N_2=N$.


\subsection{Transmission to Two Users}

When the BS adopts OMA to serve the two users in orthogonal time slots, the received signal at $u_i$ is given by $y_i = \sqrt{P_{i}}h_i x_i + n_i$.
Due to the orthogonal transmissions to $u_1$ and $u_2$, there is no interference at $u_1$ (or $u_2$) caused by $x_2$ (or $x_{1}$). As such, the SNR at $u_i$ of $x_i$ is given by 
\begin{align}
\gamma_i=\frac{P_i|h_i|^2}{\sigma_i^2}.
\end{align}
Accordingly, the decoding error probability of $x_i$ at $u_i$ for given $R_i$ is approximated by $\epsilon_i$ given in \eqref{eq:EP}. Furthermore, in OMA the effective decoding error probability $\overline{\epsilon}_i$ is $\epsilon_i$. We note that for given $R_i$ the corresponding decoding error probability $\epsilon_i$ is a function of $P_i$ (not both $P_1$ and $P_2$) due to orthogonal transmission.

\subsection{Optimization Problem in OMA}\label{sec:OMA_opt}

In OMA transmission, we have $N_1 + N_2 = N$. As such, the optimization problem given in \eqref{eq:Opt} is rewritten as
\begin{subequations}\label{eq:OptOMA}
\begin{align}
\underset{\Delta_o}{\max} ~~~&T_1 =\frac{N_1}{N}R_1(1-\epsilon_1) \label{eq:ObjectiveOMA}\\
\st~~~&T_2\geq \overline{T}_2,\label{eq:C_T_oma}\\
&N_1 P_1+ N_2 P_2 \leq N\overline{P}, \label{eq:C_Energy_OMA}
\end{align}
\end{subequations}
where $\Delta_o=\{R_1, R_2, P_1, P_2, N_1, N_2\}$ is the variable set that needs to be determined at the BS in OMA. We note that the BS has to optimize the time slots allocation on top of optimizing the transmission rates to $u_1$ and $u_2$ and the power allocation in OMA.

In order to solve \eqref{eq:OptOMA}, we first clarify that the equalities in \eqref{eq:C_T_oma} and \eqref{eq:C_Energy_OMA} are always guaranteed. This is due to the fact that for any given $R_1$, $R_2$, $N_1$, and $N_2$, the effective throughputs $T_1$ and $T_2$ are monotonically increasing functions of $P_1$ and $P_2$, respectively. Taking $T_2 = \overline{T}_2$ and $N_1 P_1+ N_2 P_2 = \overline{P}$ as constraints, we first fix the time slots allocation (i.e., fix $N_1$ and $N_2$) and then optimize $R_1$, $R_2$, $P_1$, and $P_2$ to maximize the effective throughput $T_1$ subject to the given constraints. To this end, we note that the optimal values of $R_2$, $P_1$, and $P_2$ can be efficiently determined as per $\epsilon_2$ subject to the constraints $T_2 = \overline{T}_2$ and $N_1 P_1+ N_2 P_2 = \overline{P}$ for any given $N_1$ and $N_2$, which are independent of $R_1$. We next prove that there always exists a certain coding rate $R_1$ to maximize $T_1$ and determine this optimal value.

%

\begin{theorem}\label{Theorem2}
There always exists a certain coding rate $R_1$ that maximizes $T_1$. This optimal value, denoted by $R_{1}^{\ast}$, is the unique solution to
\begin{align}
Q\left(\frac{c-R_{1}^{\ast}}{d}\right)-\frac{R_{1}^{\ast}}{d\sqrt{2\pi}}e^{-\frac{(c-R_{1}^{\ast})^2}{2d^2}}=1,
\end{align}
where $c=\log_2\left(1+\gamma_1\right)$ and
\begin{align}
d=\frac{1}{\ln2}\sqrt{\frac{1}{N_1}\left(1-\frac{1}{(1+\gamma_1)^2}\right)}.\notag
\end{align}
\begin{IEEEproof}
The proof is similar to \textit{\textbf{Theorem}~\ref{Theorem1}} given in Appendix~\ref{App:NOMA} and thus omitted here.
\end{IEEEproof}
\end{theorem}

We note that the determined optimal values of $P_1$, $P_2$, $R_1$, and $R_2$ are functions of $N_1$ and $N_2$. Finally, we are able to adopt a one-dimension numerical search method to find the optimal $N_1$ and $N_2$, which are used to determine the global optimal $P_1$, $P_2$, $R_1$, and $R_2$ together with the maximum value of $T_1$, denoted by $T_1^{\ast}$.

\section{Numerical Results}\label{sec:simulation}

In this section, we present numerical results to evaluate the performance of NOMA with finite blocklength with OMA being the benchmark. Without elsewhere stated, we set the noise power at each user to unit, i.e., $\sigma_{1}^{2}=\sigma_2^2=1$, in this section. We also define the average SNR as $\overline{\gamma} = {\overline{P}}/{\sigma_i^2}$.

\begin{figure}[t!]
    \begin{center}
        \includegraphics[height=2.45in]{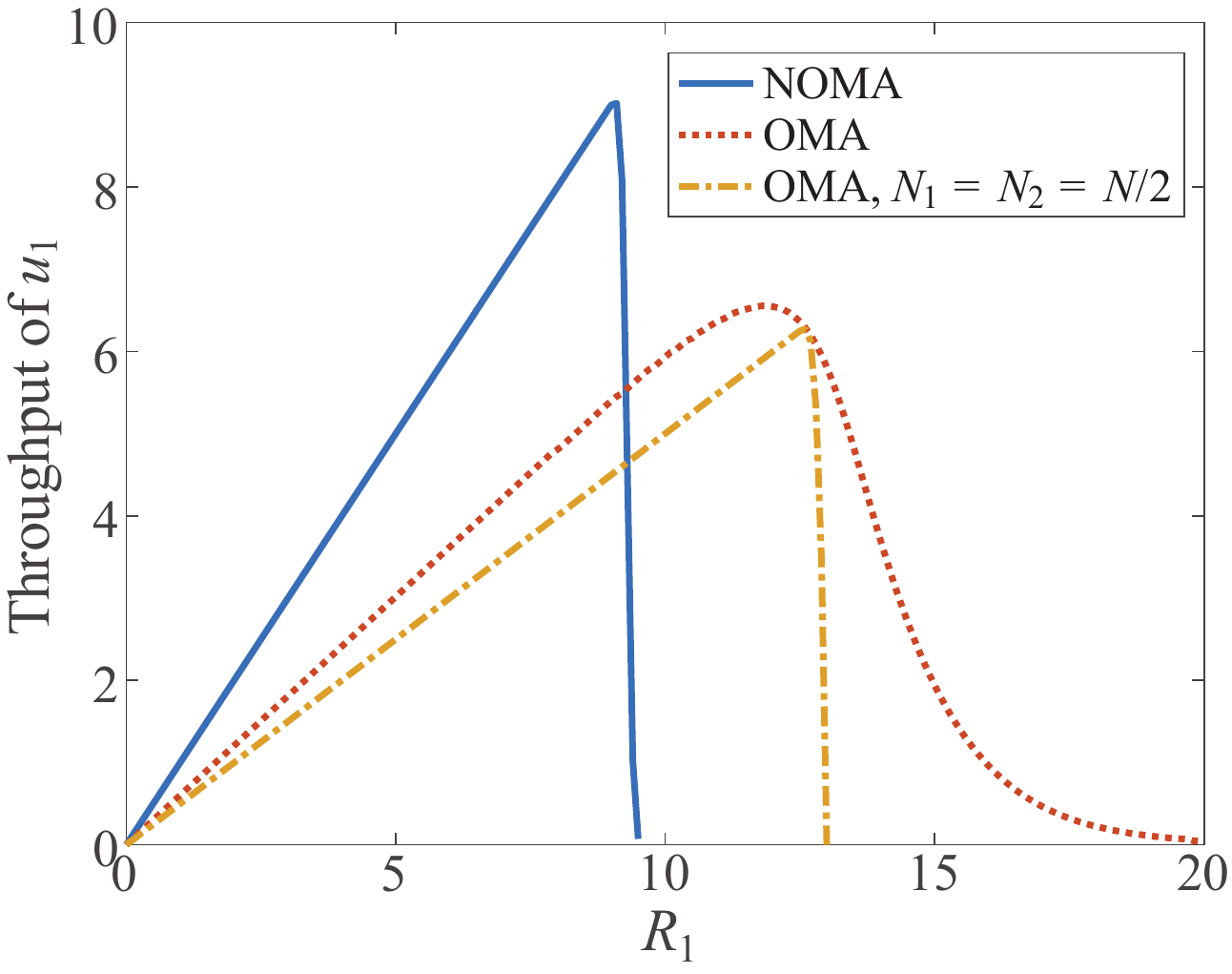}\vspace{-2ex}
        \caption{Effective throughput $T_1$ achieved by NOMA and OMA versus $R_1$ with $\overline{T}_2=3$, $\overline{\gamma}=40\dB$, $N=300$, $\left|h_{1}\right|=0.8$, and $\left|h_{2}\right|=0.1$.}
        \label{fig:NOMA_OMA_R1}
        \vspace{-4ex}
    \end{center}
\end{figure}

In Fig.~\ref{fig:NOMA_OMA_R1}, we plot the effective throughput of $u_1$ achieved by NOMA and OMA versus $R_1$, where the power allocation and $R_2$ are determined to ensure $T_2 = \overline{T}_2$ subject to $N_1 P_1 + N_2 P_2 = N \overline{P}$. In this figure, we first observe that there is an optimal value of $R_1$ that maximizes $T_1$ in both NOMA and OMA. We also observe that $T_1$ is concave with respect to $R_1$ (for fixed $N_1$ and $N_2$ in OMA), which verifies \textit{\textbf{Theorems}~\ref{Theorem1}} and~\textit{\ref{Theorem2}}. Furthermore, we observe that for the optimal $R_1$, NOMA significantly outperforms OMA by achieving a higher $T_1$, which demonstrates the advantage of NOMA in short-packet communications. This is mainly due to the fact that NOMA serves two users simultaneously and thus reduces the negative impact of the finite blocklength on the throughput. Finally, we observe that NOMA may not outperform OMA if $R_1$ is predetermined, e.g., $R_{1}=10$. This is due to the interference between the two users in NOMA but not in OMA.



\begin{figure}[t!]
    \begin{center}
        \includegraphics[height=2.45in]{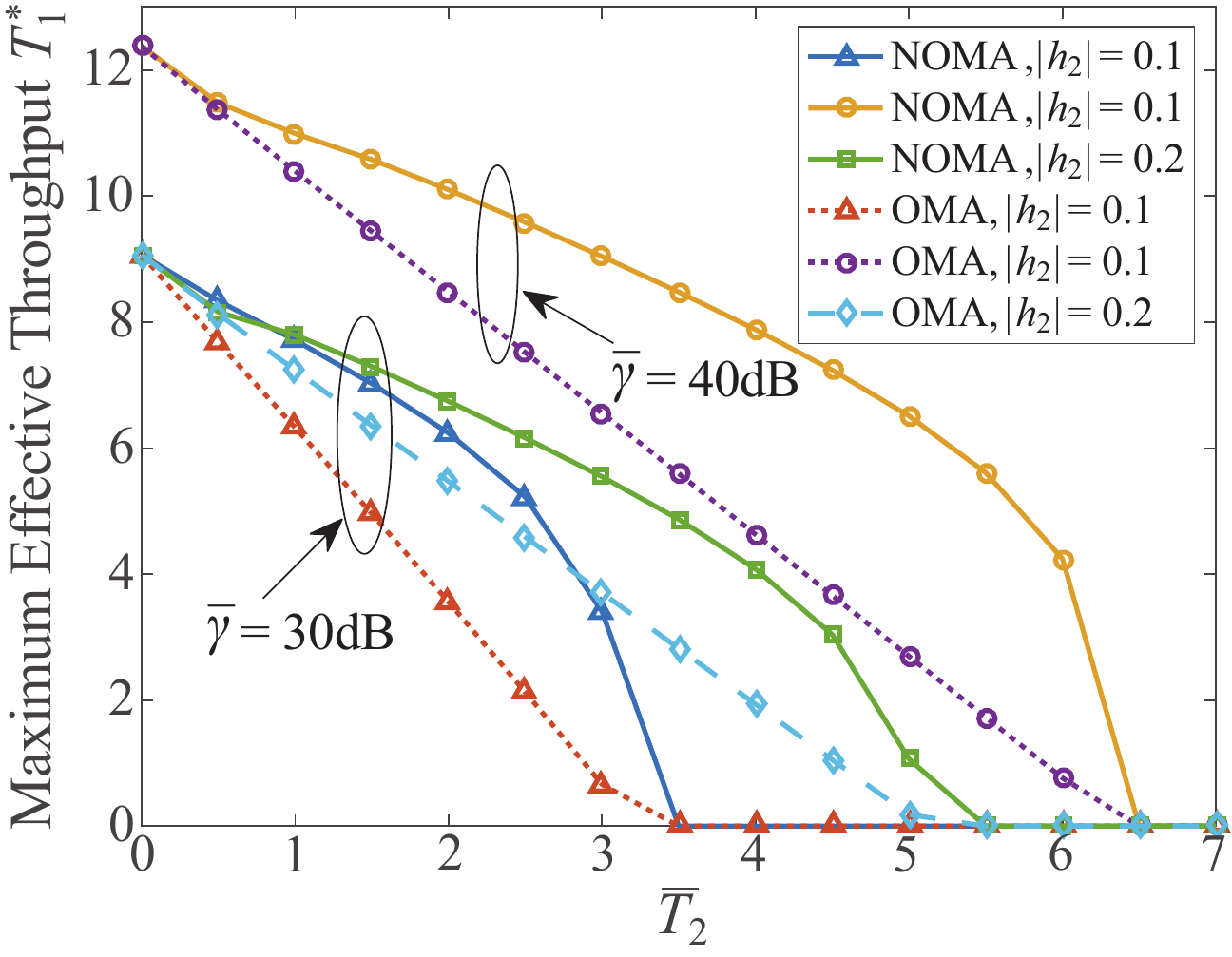}\vspace{-2ex}
        \caption{Maximum effective throughput $T_1^{\ast}$ achieved by NOMA and OMA versus  $\overline{T}_2$ with $N=300$ and $\left|h_{1}\right|=0.8$.}
        \label{fig:Throughput}
    \end{center}
    \vspace{-2ex}
\end{figure}

In Fig.~\ref{fig:Throughput}, we plot the maximum effective throughput of $u_1$, i.e., $T_1^{\ast}$, versus the constraint on the effective throughput of $u_2$, i.e., $\overline{T}_2$. First, we observe that $T_1^{\ast}$ achieved by NOMA is higher than that achieved by OMA, which is expected. Second, we observe that NOMA and OMA achieve the same maximum effective throughput when $\overline{T}_2 = 0$, since in this case the BS does not have to transmit to $u_2$ and NOMA becomes OMA. Third, we observe that the gap between $T_1^{\ast}$ achieved by NOMA and that achieved by OMA reaches the maximum point in the medium regime of $\overline{T}_2$. This demonstrates that the advantage of NOMA over OMA becomes more prominent when the effective throughput targets at $u_1$ and $u_2$ are closer to each other, which applies to most practical scenarios. Fourth, we observe that NOMA and OMA achieve the same maximum effective throughput when $\overline{T}_2$ becomes very large. This is due to the fact that when $\overline{T}_2$ is large, the constraint $T_2 \geq \overline{T}_2$ is hard to guarantee even if all the recourses are allocated to $u_2$. Fifth, we observe that when $\overline{\gamma}$ becomes higher, $T_1^{\ast}$ increases and the performance gap between NOMA and OMA increases. Sixth, we observe that the performance gap increases as the difference between $\left|h_{1}\right|$ and $\left|h_{2}\right|$ increases. In addition, we highlight that in Fig.~\ref{fig:Throughput}, the power allocation and $R_2$ are not optimal to NOMA but optimal to OMA, as discussed in Sections~\ref{sec:NOMA_opt} and~\ref{sec:OMA_opt}. Therefore, we conclude that NOMA with optimal power allocation and $R_2$ definitely achieves a superior performance compared with OMA.

\begin{figure}[t!]
    \begin{center}
        \includegraphics[height=2.45in]{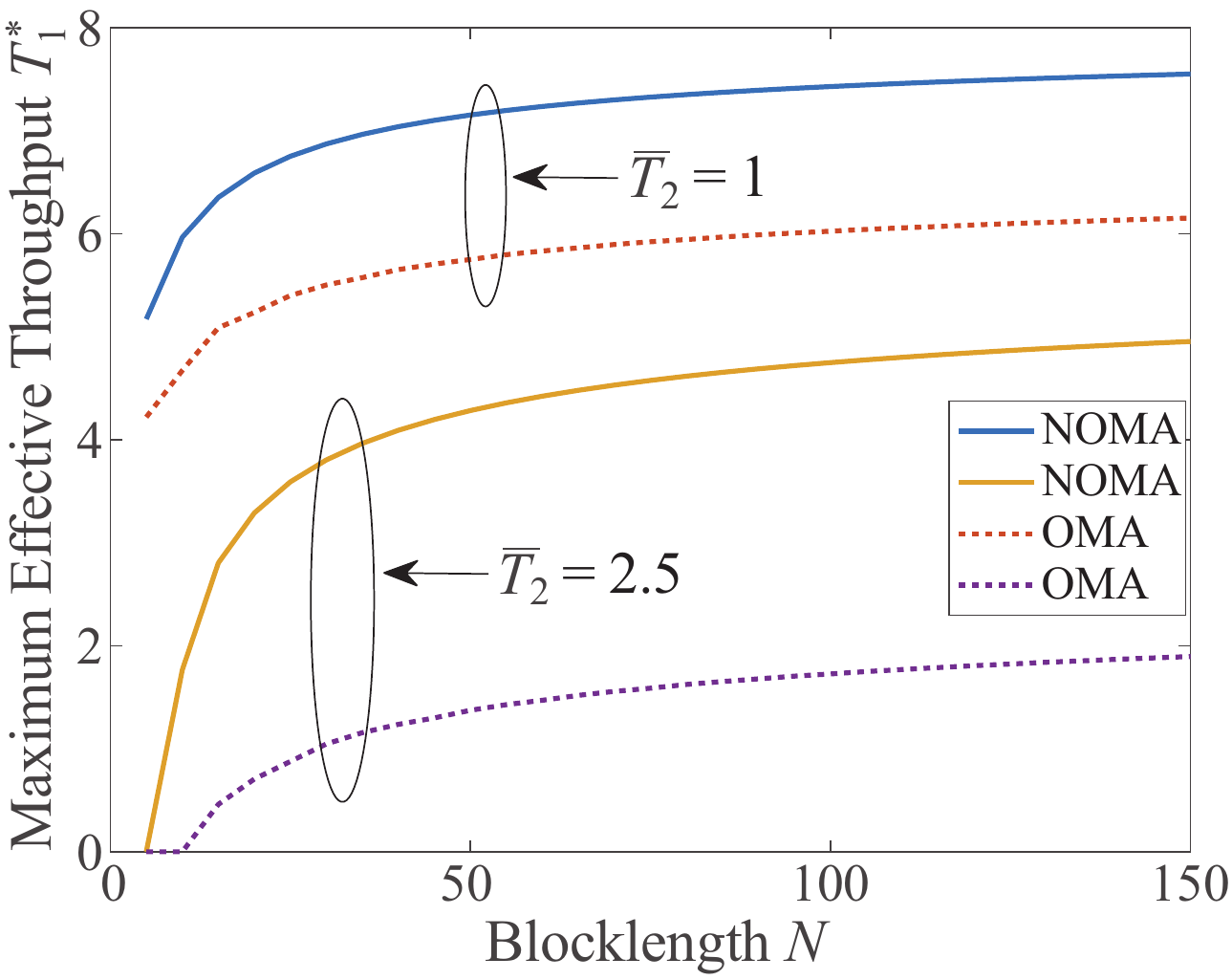}\vspace{-2ex}
        \caption{Maximum effective throughput $T_1^{\ast}$ achieved by NOMA and OMA versus blocklength $N$ with $\overline{\gamma}=30\dB$, $\left|h_{1}\right|=0.8$, and $\left|h_{2}\right|=0.1$.}
        \label{fig:Symbol_Length}
    \end{center}
    \vspace{-2ex}
\end{figure}

In Fig.~\ref{fig:Symbol_Length}, we plot the maximum effective throughput $T_1^{\ast}$ versus the blocklength $N$ for fixed $\overline{T}_2$. In this figure, we first observe that NOMA outperforms OMA regardless of $N$. We then observe that the gap between NOMA and OMA initially increases and then keeps constant as $N$ increases. Comparing the value of $N$ for fixed $T_1^{\ast}$ (i.e., $T_1^{\ast} = 5.97$) in this figure, we observe that the required blocklength of NOMA is significantly less than that of OMA (e.g., $N=10$ in NOMA while $N=85$ in OMA when $\overline{T}_2=1$). This demonstrates that NOMA can achieve a significantly lower latency than OMA for achieving the same effective throughput. Importantly, this indicates that the NOMA technique enhances the performance of short-packet communications.



\section{Conclusion}\label{sec:conclusion}

This work introduced, for the first time, NOMA in short-packet communications to achieve low latency. With OMA as a benchmark, our analytical and numerical examination demonstrated that NOMA can outperform OMA by achieving a much higher effective throughput at the user with a higher channel gain while ensuring the same effective throughput at the other user. This indicates that NOMA significantly reduces the latency in short-packet communications for achieving the same effective throughput. We further found that the performance gap between NOMA and OMA becomes more prominent as the effective throughputs at the two users become more comparable.

\begin{appendices}

\section{Proof of Lemma \ref{lemma1}}\label{App:inequality}

We now prove \textit{\textbf{Lemma}~\ref{lemma1}} by contradiction for any given $R_1$ and $R_2$ as follows:

We first assume that the current optimal power allocation $P_1^{\dag}$ and $P_2^{\dag}$, satisfying $P_1^{\dag}+P_2^{\dag}<\overline{P}$, can maximize $T_1$ (denoted by $T_1^{\dag}$) and guarantee the constraint $T_2 \geq \overline{T}_2$. We then increase $P_1^{\dag}$ and $P_2^{\dag}$ by multiplying a common scalar, $\alpha={\overline{P}}/({P_1^{\dag}+P_2^{\dag}})$, to obtain a new power allocation $P_1^{\ddag}$ and $P_2^{\ddag}$, satisfying $P_1^{\ddag}+P_2^{\ddag}=\overline{P}$. We note that $P_1^{\ddag} > P_1^{\dag}$ and $P_2^{\ddag} > P_2^{\dag}$, due to $\alpha > 1$ caused by $P_1^{\dag}+P_2^{\dag}<\overline{P}$. Following \eqref{eq:SINR_SIC},~\eqref{eq:SINR1},~\eqref{eq:SNR1}, and \eqref{eq:SINR2}, we find that $\gamma_{2}^{1},\gamma_1,\gamma'_{1}$, and $\gamma_2$ increase as $P_1^{\dag}$ and $P_2^{\dag}$ increase to $P_1^{\ddag}$ and $P_2^{\ddag}$, respectively. This leads to the fact that the corresponding decoding error probabilities $\epsilon_1$, $\epsilon'_{1}$, $\epsilon_{2}^{1}$, and $\epsilon_2$ decrease. We note that $T_2$ increases as $\epsilon_2$ decreases for a fixed $R_2$ and thus, the constraint $T_2 \geq \overline{T}_2$ is still guaranteed by $P_1^{\ddag}$ and $P_2^{\ddag}$.

Following \eqref{eq:Average_EP1} and noting $\epsilon_1 < \epsilon'_{1}$ due to $\gamma_1 > \gamma'_{1}$, we find that $\overline{\epsilon}_1$ is a monotonically increasing function of $\epsilon_1$,  $\epsilon'_{1}$, and $\epsilon_{2}^{1}$, respectively. As such, $\overline{\epsilon}_1$ decreases as $P_1^{\dag}$ and $P_2^{\dag}$ increase to $P_1^{\ddag}$ and $P_2^{\ddag}$, respectively. It follows that $T_1$ increases from $T_1^{\dag}$ to $T_1^{\ddag}$ (i.e., $T_1^{\dag} < T_1^{\ddag}$). Notably, this contradicts to the original assumption that $T_1^{\dag}$ is the maximum value of $T_1$ achieved by $P_1^{\dag}$ and $P_2^{\dag}$. Therefore, we conclude that $P_1+P_2=\overline{P}$ is always guaranteed in the optimization problem given in \eqref{eq:Opt1}.

\section{Proof of Theorem \ref{Theorem1}}\label{App:NOMA}

Following \eqref{eq:Throughput}, we rewrite $T_1$ as a function of $R_1$ as
\begin{align}\label{eq:f(R1)}
\mathcal{T}\left(R_{1}\right)=&R_1\left(1-Q\left(\frac{a-R_1}{b}\right)\right)(1-\epsilon_{2}^{1})\notag\\
&+R_1\left(1-Q\left(\frac{a'-R_1}{b'}\right)\right)\epsilon_{2}^{1}.
\end{align}
To determine the optimal value of $R_1$ that maximizes $\mathcal{T}\left(R_{1}\right)$, we next examine the monotonicity and concavity of $\mathcal{T}\left(R_{1}\right)$ with respect to $R_1$. To this end, we derive the first and second derivatives of $\mathcal{T}\left(R_{1}\right)$ with respect to $R_1$ as follows:

Based on the differentiation of a definite integral with respect to a parameter~\cite{Table_2007}, the first derivative of $\mathcal{T}\left(R_{1}\right)$ with respect to $R_1$ is derived as
\begin{align}\label{eq:f'R1}
\mathcal{T}'\left(R_{1}\right)=& (1-\epsilon_{2}^{1})\left(1-Q\left(\frac{a-R_1}{b}\right)-\frac{R_{1}\tau}{b}\right)\notag\\
&+\epsilon_{2}^{1}\left(1-Q\left(\frac{a'-R_1}{b'}\right)-\frac{R_{1}\tau'}{b'}\right),
\end{align}
where $\tau=e^{-\frac{(a-R_1)^2}{2b^2}}/\sqrt{2\pi}$ and $\tau'=e^{-\frac{(a'-R_1)^2}{2b'^2}}/\sqrt{2\pi}$.

Similarly, the second derivative of $\mathcal{T}\left(R_{1}\right)$ with respect to $R_1$ is derived as
\begin{align}\label{eq:second_f}
\mathcal{T}''\left(R_{1}\right)=&
(1-\epsilon_{2}^{1})\left(-\frac{2\tau}{b}-\frac{\left(a-R_1\right)R_{1}\tau}{b^3}\right)\notag\\
&+\epsilon_{2}^{1}\left(-\frac{2\tau'}{b'}-\frac{\left(a'-R_1\right)R_{1}\tau'}{b'^3}\right).
\end{align}
We note that $a\geq 0$, $a'\geq 0$, $b\geq 0$, and $b'\geq 0$ must hold based on their respective values. Thus, we have $\mathcal{T}'(0)>0$, due to $0\leq Q(\frac{a}{b})\leq 0.5$ and $0\leq Q(\frac{a'}{b'})\leq 0.5$, and have $\mathcal{T}''(0) < 0$. When $0\leq R_1\leq a'$, following \eqref{eq:second_f} we find that $\mathcal{T}''(R_1)$ keeps negative. Thus, $\mathcal{T}'(R_1)$ keeps decreasing within $0\leq R_1\leq a'$.

When $R_1>a'$, we find that the error probability of $u_1$ when SIC is not guaranteed is larger than $0.5$, i.e., $\epsilon'_{1}>0.5$, which leads to zero throughput. As such, the second item in the right-hand side of \eqref{eq:f(R1)} is negligible when $R_1>a'$. Due to this, we find that $\mathcal{T}''(R_1)$ still keeps negative and $\mathcal{T}'(R_1)$ keeps decreasing as $R_1$ increases to $a$. Furthermore, when $R_1$ approaches infinity, both $Q\left(a-R_1)/b\right)$ and $Q\left((a'-R_1)/b'\right)$ approach one, whereas both $\frac{R_1}{b\sqrt{2\pi}}\exp\left(-\frac{(a-R_1)^2}{2b^2}\right)$ and $\frac{R_1}{b'\sqrt{2\pi}}\exp\left(-\frac{(a'-R_1)^2}{2b'^2}\right)$ approach zero, as per the L'Hospital's rule. Thus, we obtain $\mathcal{T}'(R_1) \rightarrow 0$ as $R_1 \rightarrow \infty$.

We next denote the values of $R_1$ which satisfy $\mathcal{T}'(R_1)=0$ and $\mathcal{T}''(R_1)=0$ by $R_1^{\dag}$ and $R_1^{\ddag}$, respectively. We note that $\mathcal{T}'(R_1)$ decreases from a positive value to $\mathcal{T}'(R_1^{\ddag})$ as $R_1$ increases from $0$ to $R_1^{\ddag}$. Since $\mathcal{T}'(R_1)$ is a continuous function and $\mathcal{T}''(R_1)$ keeps positive when $R_1\geq R_1^{\ddag}$, $\mathcal{T}'(R_1)$ increases from $\mathcal{T}'(R_1^{\ddag})$ to $0$ as $R_1$ increases from $R_1^{\ddag}$ to infinity, due to $\mathcal{T}'(\infty) \rightarrow 0$. This indicates $\mathcal{T}'(R_1^{\ddag}) \leq 0$ and $R_1^{\dag} \leq R_1^{\ddag}$. Therefore, we conclude that when $R_1\leq R_1^{\ddag}$, $\mathcal{T}\left(R_{1}\right)$ is strictly concave with respect to $R_1$. It follows that the optimal coding rate $R_{1}^{\ast}$ is $R_1^{\dag}$, i.e., $R_1^{\ast} = R_1^{\dag}$, which completes the proof.

\end{appendices}

\renewcommand\refname{References}

\bibliographystyle{IEEEtran}
{\footnotesize\bibliography{IEEEabrv,NOMA_FBL}}

\end{document}